\documentclass[journal]{IEEEtran}
\usepackage{hyperref,listings}
\usepackage[table,xcdraw]{xcolor}
\usepackage{graphicx}

\title{Toward a Blockchain-based Platform to Manage Cybersecurity Certification of IoT devices}

\author{Ricardo Neisse,  Jos\'e L. Hern\'andez-Ramos, Sara N. Matheu, Gianmarco Baldini and  Antonio Skarmeta,~\IEEEmembership{Member,~IEEE} \thanks{

Ricardo Neisse, Jos\'e L. Hern\'andez-Ramos and Gianmarco Baldini are with European Commission, Joint Research Centre, Ispra 21027, Italy; emails: \{ricardo.neisse, jose-luis.hernandez-ramos, gianmarco.baldini\}@ec.europa.eu

Sara N. Matheu and Antonio Skarmeta are with University of Murcia, Department of Information and Communications Engineering; emails: \{saranieves.matheu, skarmeta\}@um.es}
}

\newcolumntype{M}[1]{>{\centering\arraybackslash}m{#1}}
\begin{document}

\markboth{}{A blockchain-based platform to track the security assessment of IoT devices}
\maketitle

\begin{abstract}
The goal of this paper is to propose a blockchain-based platform to enhance transparency and traceability of cybersecurity certification information motivated by the recently adopted EU Cybersecurity Act.
The proposed platform is generic and intended to support the trusted exchange of cybersecurity certification information for any electronic product, service, or process.
However, for the purposes of this paper, we focus on the case study of the cybersecurity certification of IoT devices, which are explicitly referenced in the recently adopted Cybersecurity Act as one of the main domains where it is highlighted the need for an increased level of trust.
\end{abstract}


\section{Introduction}
\label{sec:intro}

The advent of the so-called Internet of Things (IoT) supports an increasing integration of the physical and digital world where security and privacy aspects represent significant concerns for companies, institutions and end users due to the growing use of digital identity and  processing of personal data. Through this trend toward hyperconnectivity, security issues may not be only associated to a single device but they can also extend their negative impacts to other devices or systems. Therefore, there is a need to develop mechanisms to evaluate and assess the security level provided by IoT devices and technologies, as highlighted in \cite{khan2018iot} and \cite{kouicem2018internet}. In this way potential cybersecurity vulnerabilities and attacks can be detected and mitigated to create more secure and trustworthy IoT-enabled environments. 

As a legal umbrella for security assessment, security certification has recently emerged as a key tool in establishing a common framework to increase trust in a digital society. In this direction, the realization of a cybersecurity certification framework is nowadays one of the main EU priorities in the field of cybersecurity. Indeed, the European Commission (EC) adopted in March 2019 \cite{cybersecurity_act_EP} the so-called ``Cybersecurity Act'' with the goal of strengthening the role of the European Union Agency for Network and Information Security (ENISA), and to establish such a framework at EU level that allows the recognition of certified devices in all Member States.

Related to the development of this framework, other recent efforts have been initiated to identify the needs and challenges to be addressed in the coming years. In particular, the European Cyber Security Organization (ECSO) has analyzed the challenges of the industry in the field of cybersecurity certification, and the set of current alternatives that can be used as inputs for the definition of a meta-certification scheme \cite{ecso_meta-scheme_2017}.
These issues are exacerbated in the IoT context \cite{voas2018iot} for many reasons including the wide deployment of IoT technologies in the market and its distribution in many different applications and infrastructures, the dynamic nature, scale and pervasiveness of IoT technologies, and the demand for a common platform to keep track on the certification results of IoT devices and systems. This platform must be flexible enough to support the variety of different stakeholders in the IoT market, the different domains where IoT devices can be deployed, and their different versions and configurations.

While an exhaustive collection of the different challenges for the definition of a cybersecurity certification framework can be found in our previous works \cite{matheu2019risk} \cite{matheu2019toward}, we focus on two main aspects that need to be addressed for an effective certification approach in the IoT context.
On the one hand, the information in the cybersecurity certificate (as a result of a successful certification process) should include pertinent data for a secure deployment of a device without the involvement of end users. This is particularly relevant in IoT, since non-expert users will use such devices in their everyday lives. To address these aspects, we propose the use of the Manufacturer Usage Description (MUD) \cite{mudrfc} to be part of the certificate of a certain device or system. MUD is a recent IETF standard that provides a mechanism for devices to signal to the network the type of access and network functionality they require. Therefore, network components can adapt their behavior to enforce the access control preferences that are specified in a device's MUD file.
On the other hand, we propose the use of blockchain to realize the mentioned platform, in order to provide a unified view of the security level of an IoT device throughout its lifecycle. This information could be provided by different sources (e.g., different Member States) through a common EU certification scheme. While the use of blockchain has been widely considered in the context of IoT \cite{fernandez2018review} addressing challenges in different use cases, to the best of our knowledge, this is the first work proposing a blockchain-based platform to track the security certification of IoT devices. The development of such blockchain-based platform to share cybersecurity certificates represents a complementary effort to the development of the mentioned framework, and could improve trust and transparency aspects in the digital era. 

Even though we are the first proposing platform to track security assessment information about IoT devices, there similar initiatives in other areas that are worth mentioning.
The platform proposed in this paper is similar in functionality to the European Database on Medical Devices (EUDAMED)\footnote{\url{http://ec.europa.eu/idabc/en/document/2256/5637.html}} initiative, which aimed specifically at providing a repository of information about medical devices.
EUDAMED includes a registry of manufacturers, representatives, devices, certificate lifecycle, and vigilance information including incidents or near-incidents that occurred during the use of a medical device.
The main difference between EUDAMED and our proposal in this paper is the use of a decentralized blockchain-based approach where the stakeholders can jointly manage the cybersecurity certification information.
Blockchain technology is currently being considered in the management of official records (e.g. academic or professional licenses) in the BlockCerts open standard\footnote{https://www.blockcerts.org} using a \emph{verifiable claim} concept.
This concept could be used as a building block in our platform in future production implementations.

The structure of the paper is following: Section \ref{sec:certification-background} provides an overview of the regulatory landscape related to cybersecurity certification in the EU. Then Section \ref{sec:blockchain-background} describes the basics of blockchain as the baseline for the platform proposed in Section \ref{sec:blockchain}. Furthermore, Section \ref{sec:cert_iot} proposes the inclusion of additional information to be embedded in the security certificates to foster a secure deployment of IoT devices. A detailed description of our proof-of-concept implementation is given in Section \ref{sec:implementation}, and Section \ref{sec:conclusions} concludes the paper with an outlook about our future work in this area.

\section{Cybersecurity Certification and EU Regulatory Landscape}
\label{sec:certification-background}

Cybersecurity certification represents an essential instrument to build a more trustworthy hyper-connected digital landscape.
Towards the realization of an EU cybersecurity certification framework, the European Parliament adopted in March 2019 \cite{cybersecurity_act_EP} the Cybersecurity Act.
This regulation proposes a common umbrella for cybersecurity certification in the EU to address the current market fragmentation through a harmonized approach, with a view to creating a digital single market for ICT products, services and processes. 
This new regulation also establishes a roadmap for the development of an EU framework for cybersecurity certification, with the end goal of boosting the cybersecurity of digital products and services in Europe. The adopted regulation identifies cybersecurity certification as a key component to increase confidence in digital solutions by defining such a framework. This effort will further reduce the development of new (potentially conflicting or overlapping) schemes, in order to foster the sustainable development of the Digital Single Market.

The cybersecurity certification framework proposed by the Cybersecurity Act is intended to embrace different cybersecurity certification schemes. Therefore, the framework should be flexible enough to accommodate different schemes that could be focused on specific domains. Indeed, the regulation states 
that the content and the format of the European cybersecurity certificates is one of the elements to be included in the schemes.
However, the lack of an agreed format and content of the resulting certificate could lead to a disharmonized coexistence of EU schemes.
In this direction, the ECSO proposes to include the certificate itself, and the certification report in a European Cyber Security Certificate (ECSC). The suggested content of the ECSC includes a label for simplified understanding of the certification results, which is especially relevant for end users. Furthermore, it is intended to include some attributes related to the process, such as the Generalized Protection Profile (GPP) and Generalized Security Target (GST), which represent concepts used in the Common Criteria standards \cite{ccra_common_2017}.

Indeed, as already mentioned, the fragmentation of cybersecurity certification schemes at EU level represents a significant challenge to be addressed by this framework.
There is a need to foster cooperation among EU Member States to achieve a harmonized vision of cybersecurity risks in our society. 
In this sense, one of the main existing efforts is represented by the Senior Officials Group - Information Systems Security (SOG-IS) Mutual Recognition Agreement (MRA). SOG-IS MRA was created to foster cooperation and mutual recognition through the standardization of CC protection profiles.

This aspect is also highlighted through the survey carried out by ENISA in 2017 \cite{enisa_report} regarding the development of a cybersecurity certification framework at EU level. According to the results of this survey, the responses of different industry representatives emphasize the need to improve current processes and tools for cybersecurity certification \cite{mitrakas2018emerging}.
In addition to the issues related to the high cost and long duration of the certification process, the results highlight the concerns regarding the need for mutual recognition at EU level for products and services. Another key aspect is the need to have self-declaration schemes based on a self-assessment process to cope with a fast moving market. Indeed, this is also contemplated by the Cybersecurity Act \cite{cybersecurity_act_EP}, 
which specifies that the manufacturer or provider shall assume the responsibility for the compliance of the product in case of using a self-assessment process. An additional point was stressed regarding the need for certification and labelling processes in the IoT domain, due to its ubiquitous nature and required interoperability across different platforms. 

According to the Cybersecurity Act \cite{cybersecurity_act_EP}, as the result of a successful evaluation of an ICT product, service, or process, a cybersecurity certificate is issued depending on a certain assurance level.
%
This level represents a basis for confidence that an ICT product, service or process meets the security requirements of a certain scheme. The value of the assurance level could be ``basic'', ``substantial'' and ``high'', and it aims to provide the corresponding rigor and depth of the evaluation. It represents one of the elements to be defined by each European cybersecurity certification scheme.
%
Furthermore, the regulation also defines supplementary cybersecurity information to be provided by the manufacturer or provider in electronic format, including guidelines and recommendations about the use of the product or service to end users, contact information of the manufacturer or provider, as well as a reference to an online repository listing publicly disclosed vulnerabilities.
In this direction, as will be described in Section \ref{sec:cert_iot}, we consider the inclusion of guidelines and recommendations by using the recent MUD standard, which defines a format to define the intended use of a specific device or system. Furthermore, the use of blockchain in our architecture presented in Section \ref{sec:blockchain} is intended to support the mentioned online repository.

The adopted regulation also identifies the following set of roles to be considered in the cybersecurity certification ecosystem:
\begin{itemize}
    \item National Cybersecurity Certification Authority (NCCA): among other aspects, this is the entity responsible for the supervision, enforcement, monitoring of the compliance of ICT products, services or processes with the requirements of the corresponding certificates. They also monitor and enforce the obligations of manufacturers established in their territories, as well to assist the national accreditation bodies in the supervision of conformity assessment bodies;
    \item Conformity Assessment Body (CAB): it is in charge of performing a conformity assessment process for evaluating whether specified requirements of a certain ICT product, service, or process have been fulfilled. They are accredited by a national accreditation body under certain requirements A CAB is intended to issue European cybersecurity certificates referring to assurance level “basic” or “substantial”. In the case of level “high”, they must be issue by a NCCA, or a CAB under specific conditions; 
    \item European Cybersecurity Certification Group (ECCG): the ECCG is intended to be composed by representatives of different NCCAs to advise and assist the EC in cybersecurity certification aspects. 
\end{itemize}

Based on these roles, we propose a blockchain-based platform, which is intended to foster a trusted and transparent sharing process of the cybersecurity certification results of devices and products. Details of such a platform are provided in Section \ref{sec:blockchain}

\section{Blockchain-Based Distributed Ledgers}
\label{sec:blockchain-background}

In this section we describe the Ethereum Virtual Machine (EVM) \cite{ethereum}, which is a Blockchain-based Distributed Ledger Platform used as a building block in the solution proposed in this paper.
The EVM is currently used in a public blockchain instance where anyone with a computer connected to the Internet can submit and validate transaction blocks.
The EVM addresses many challenges including synchronization issues, security, and soundness of the distributed protocol.
In this section we limit ourselves to the high-level description of the EVM from a user perspective including design choices for systems using the EVM as a building block without presenting details about how these challenges were addressed.

Blockchain implementations based on the EVM can be primarily used to keep a decentralized record of transactions that are associated to a virtual currency (ETH), which are essentially transfers of balances from a source to a target user account.
In addition to virtual currency transactions, the EVM can also be used to deploy custom software in the blockchain, usually referred to as smart contracts, or to trigger the execution of functions in the deployed contracts.
An EVM blockchain instance can be defined as a distributed database of blocks of transactions managed by nodes that do not necessarily trust each other and do not share a common trusted third party.

In the EVM there are three types of nodes: clients, full nodes that act as observers and store the complete blockchain, and full nodes that act as transaction validators and block builders (a.k.a. miner nodes).
Client nodes submit transactions to miner nodes, which are responsible for assembling blocks containing a set of validated transactions, for which a cryptographic puzzle must be solved as a proof-of-work consensus.
The first miner node to successfully solve the puzzle broadcasts the block of transactions including the answer to the puzzle to all other full nodes, which then verify/accept the block and start working on the next block.
As a reward, the miner node receives a fixes amount of virtual currency and also collects fees for the transactions included in the block.
Each block of transactions include a reference to the previous block, in order to ensure the integrity and immutability properties since any change to a previous mined block would invalidate the references on all blocks mined afterwards.
The first block for a specific blockchain instance must be pre-agreed between the miner nodes and is called the genesis block.

Client nodes must sign all transactions submitted using a randomly generated private key, which is used to derive their public key and EVM public address identifier.
The same type of identifier is also used for smart contracts, however, contracts do not hold a private/public key pair since they are stored in the public blocks.
Smart contracts are simply assigned a random identifier when they are deployed, which is used for future reference.
EVM public addresses are not verified by any means and clients are free to generate new random addresses any time they wish to prevent linkability of their transactions.

Smart contracts in the EVM are stateful turing-complete programs implemented using a custom language named Solidity\footnote{\url{https://ethereum.github.io/browser-solidity/}}.
From a practical perspective smart contracts in the EVM are very simple and can be implemented using a very restrictive number of instructions, data structures, and events.
Therefore, in most cases, smart contracts simply act as simple registers of verified information that is made available on databases out of the blockchain.

The EVM by design targets the implementation of public blockchain instances, however, other alternative deployments are possibly namely a semi-public consortium or private.
The choice of the type of deployment impacts on the access control, auditability, and censorship resistance properties since it regulates who is allowed to submit transaction, validate transactions/blocks, and observe full log of transactions \cite{VitalikPublicPrivateBlockchains}.
In this paper we propose to use a public blockchain and we propose identity management smart contracts to be used in order to allow some level of assurance on the identification of the different stakeholders.
For example, it is important to be able to reliably identify device owners, device manufacturers, EU Member State authorities, and Conformity Assessment Bodies.
Details about our approach are introduced in the following section.

\section{Using Blockchain for Trusted Sharing of Cybersecurity Assessment Information}
\label{sec:blockchain}

In this section, we present the design of a blockchain-based platform for trusted sharing of cybersecurity certification information.
This platform consists of smart contracts for identity management and device registry deployed in a blockchain-based virtual machine.
The platform is not specific for any particular cybersecurity certification information and can be used without major changes for other types of information as needed.

The primary goal of this platform is to be used at EU level when device manufacturers and conformity assessment bodies are distributed across different Member State countries.
For example, if a device manufacturer located in Spain wants to sell a device to consumers in Germany that was certified by a Conformity Assessment Body located in Italy there should be a way of ensuring reliable identification and exchange of cybersecurity assurance information.
A secondary application of our platform is to track the supply chain of the device parts and software components also considering manufacturer and assessment bodies outside of the EU.

\begin{figure*}[!ht]
\centering
 \includegraphics[width=0.96\textwidth]{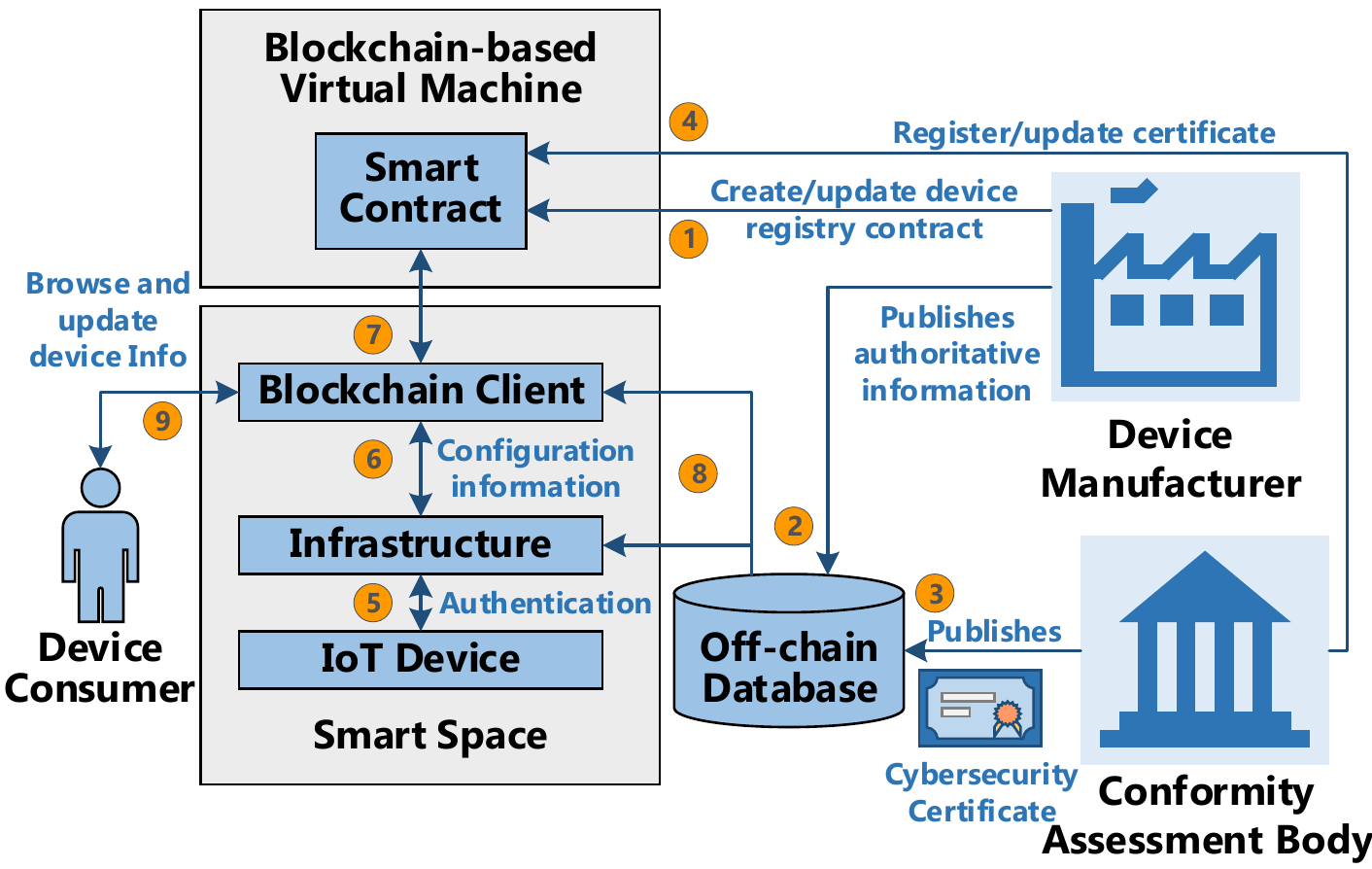}
\caption{Platform architecture}
 \label{fig:architecture}
\end{figure*}

Figure \ref{fig:architecture} presents the architecture elements including a sequence of activities.
The first step is for the manufacturer of the IoT device to create a smart contract that acts as a registry for authoritative device information that is published in an off-chain database (activities 1 and 2).
This smart contract contains various information including, for example, the manufacturer name, contact information, identity certificate, device type, device id, last firmware version and hash/fingerprint, and a MUD file describing the typical network interactions (see Subsection \ref{sec:mud}).
A Conformity Assessment Body is able to access the smart contract and update it by adding information about a cybersecurity assessment certificate, which is also published in an off-chain database (activities 3 and 4).
When an IoT device is deployed in a Smart Space, the local infrastructure, after authenticating the device (activity 5), can then retrieve any relevant device information using a Blockchain Client that is kept in sync with the current state of the device registry Smart Contract (activities 6 and 7).
The Blockchain Client or local Infrastructure may then also retrieve any information from the off-chain database as needed to setup the IoT device or to inform the Device Consumer about the device features, capabilities, and cybersecurity certificates and labels.
The Blockchain Client may also act as a point of entry for any information the Device Consumer may want to publish about the device, for example, any user experience issues or vulnerability information.

In the architecture presented in Figure \ref{fig:architecture}, no precise details are presented about the identification and authentication of all stakeholders. However, this is an important aspect of the solution we propose.
Device Manufacturers and Certification Authorities are expected to be dispersed over all EU Member States, following a common standard for publishing information about IoT Devices and Cybersecurity Certificates (See Section \ref{sec:certification-background}).
In order to ensure a global EU registry of manufacturers and certification authorities we propose a hierarchical approach where an EU Identification Service is responsible for issuing certificates for each Member State Certification Authority, which is in charge of issuing certificates for their national device manufacturers and conformity assessment bodies.
The Device Manufacturer is then in charge of providing identification information for the IoT Device and for verified device consumers, meaning that these are consumers that in fact own a specific IoT device type/model.
By registering the identification information using a newly generated blockchain identity, the actual identity of the device consumer is anonymized meaning that, for each owned device, the device consumer uses an independent identity.
By adopting this anonymization approach it is not possible to derive from the blockchain all devices owned by a specific Consumer even if multiple device manufacturers collude.

\begin{figure}[!ht]
\centering
 \includegraphics[width=0.48\textwidth]{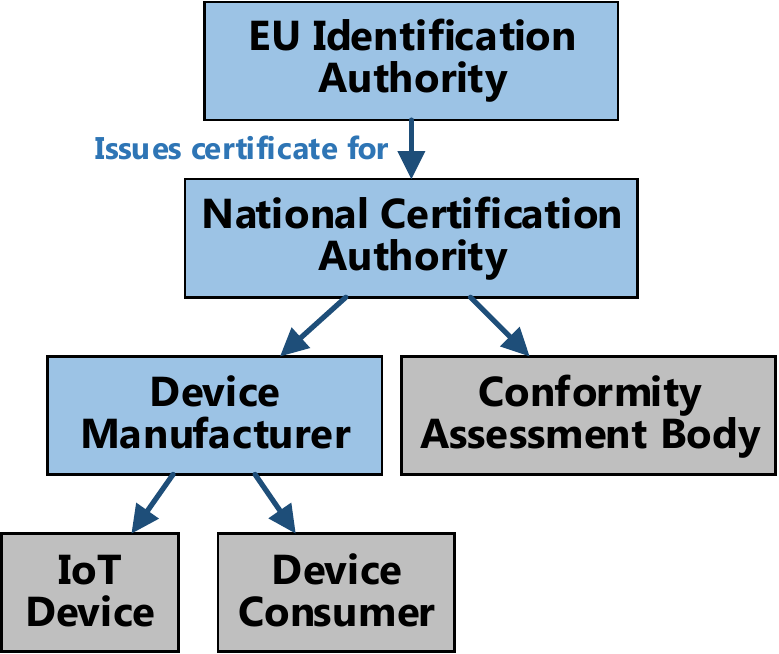}
\caption{Identification hierarchy}
 \label{fig:identification}
\end{figure}

The platform proposed in this paper relies on public blockchain implementation, however, considering the transaction fees and the typical limited transaction throughput, this may not be the optimal approach.
Our focus on this paper is on the design of the platform and a proof-of-concept prototype implementation. Practical deployment design choices including the possible adoption of a semi-public/consortium blockchain implementation are part of our future work.

\section{Enriching cybersecurity certification in IoT} \label{sec:cert_iot}

As a result of a successful cybersecurity evaluation, a cybersecurity certificate should be issued by the corresponding authority. The regulation defines this term as a ``\textit{document issued by a relevant body, attesting that a given ICT product, ICT service or ICT process has been evaluated for compliance with specific security requirements laid down in a European cybersecurity certification scheme}''. The content to be included in this certificate should help to make informed choices taking into account the expected technical level of end users.

Indeed, in the IoT context, non-expert users should be able to understand the security level provided by a certain ICT component. Furthermore, they should be entitled to access the information related to the certification scheme, including the assurance level and associated risks, as well as regarding the expected updates or patches for that component. Consequently, there is a need to define a transparent approach to empower users with the ability to access the information about the cybersecurity of their ICT products, services, and processes. In addition, the regulation also proposes guidance on action or setting that the end user can implement to maintain or increase the cybersecurity of a certain component. Again, such guidelines must consider the involvement of non-expert users, who could be opposed to adopt them. 

Beyond these general aspects, below we describe some potential aspects to be part of the European cybersecurity certificate by considering the specifics of the IoT context. These elements are intended to ease a secure and automated deployment of IoT devices.

\subsection{Manufacturer Usage Description (MUD)} \label{sec:mud}
The goal of the Manufacturer Usage Description (MUD) standard \cite{mudrfc} is to define a format and message exchange for end devices to signal access control lists to the network where they are deployed. The specification of these access control lists are specified by manufacturers in a MUD file by using a Yet Another Next Generation (YANG) model \cite{bjorklund2016yang} to describe the intended behavior of a certain device. Consequently, the deployment of MUD files could reduce the threat and vulnerability surface of a device to the communications defined by the manufacturer. Furthermore, it gives a certain degree of extensibility, so that manufacturer can express other device requirements. 

MUD could cover a gap in the scope of security certification, in order to support a secure and automated deployment of IoT devices. In particular, according to the Cybersecurity Act (Article 55), the manufacturer or provider shall make publicly available (among other things) \textit{``guidance and recommendations to assist end users with the secure configuration, installation, deployment, operation and maintenance of the ICT products or ICT services;''}. In this direction, the use of MUD files could also help to reduce the users’ involvement when installing and deploying new IoT devices. The MUD standard is also strongly considered by the NIST in recent reports to mitigate network-based attacks in IoT \cite{dodson2019securing}. 

\subsection{Vulnerability Disclosure}

Throughout their lifecycle, new potential vulnerabilities could be detected in a certain device or product. In this direction, a coordinated vulnerability disclosure strategy could increase users’ trust in the digital era, through the cooperation between manufacturers, governments and organizations. Indeed, according the Cybersecurity Act, 
such cooperation has been proven to significantly increase both the rate of discovery and the remedy of vulnerabilities. Furthermore, it states certain conditions for the process, in which the manufacturer or provider has the ``opportunity to diagnose and remedy the vulnerability before detailed vulnerability information is disclosed to third parties or to the public''

Aligned with the need for ``facilitating the access to better-structured information on cybersecurity risks and possible remedies'', 
the use of a blockchain-based platform could help to track the state of a newly disclosed vulnerability. For instance, a certain organization could add a hash of the vulnerability to the blockchain that is encrypted to be only accessible to the manufacturer. Then, after a certain period, such organization could report the status of the vulnerability to the corresponding NCCA.

A related aspect with the vulnerability disclosure process is the lack of a vulnerability database in the EU. This database could consider the existing National Vulnerability Database (NVD) \cite{zhang2015predicting} in the U.S. We believe that such EU database could be integrated as part of the proposed blockchain-based platform to foster the cooperation and coordination among Member States through a centralized point to share products’ vulnerabilities that are marketed at EU level. 

\subsection{System modeling and testing processes}

As part of the cybersecurity information, a detailed structural and behavior models of IoT components \cite{neisse_seckit_2015} could help to come up with a more clear understanding about potential security risks.
This information could be based on the use of Model-Based Testing (MBT) \cite{utting2010practical}, which we have been considered in our previous works for the development of a risk-based security testing and assessment methodology for IoT \cite{matheu2019toward}. Indeed, such models are used for the generation of tests to be executed in the device. Then, the results of these tests (test report) are employed for estimating the risk of potential vulnerabilities.

In this case, the use of blockchain could also help to represent the relationship between different devices or components; indeed, the security level of a certain device could affect to the security provided by a whole system.
Thus, blockchain could support the tracking of the components' security level, as well as the relationship among them. An additional aspect is the inclusion of test results, which are employed to evaluate and certificate the security level of a certain component. While testing processes are already considered by the Cybersecurity Act 
(e.g., through penetration testing), there is a lack of standard mechanisms to represent the test report (and the tests themselves) in a readable-machine way, so that such tests can be re-executed and analyzed throughout the devices' lifecycle.

\section{Case Study and Implementation}
\label{sec:implementation}

\begin{figure*}[!ht]
\centering
 \includegraphics[width=.65\textwidth]{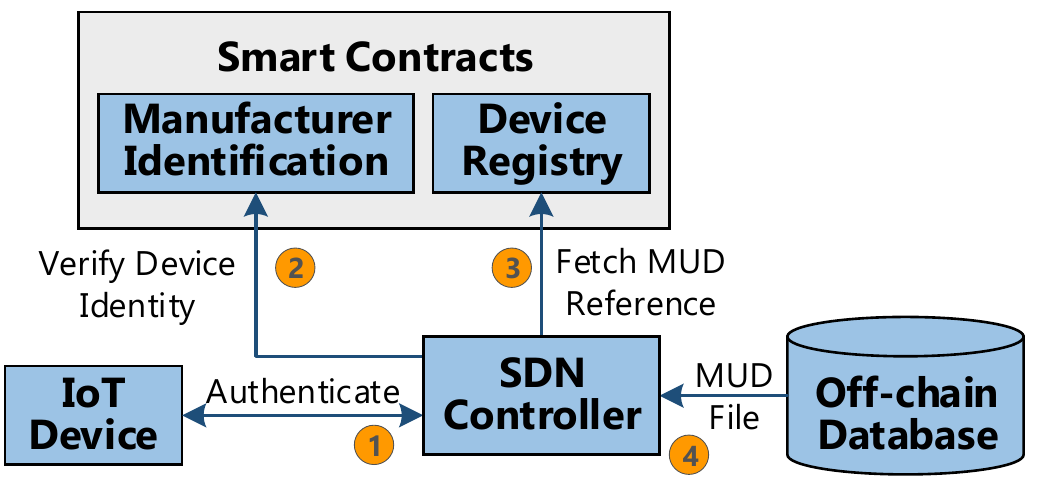}
\caption{Implementation scenario}
 \label{fig:implementation}
\end{figure*}

In this section we present a proof-of-concept implementation of our blockchain-based platform described in the previous section.
As a case study we consider the retrieval of a MUD file reference from the device registry smart 
contract by a Software Defined Network (SDN) controller.
A diagram of the implementation scenario is presented in Figure \ref{fig:implementation}.
The SDN controller authenticates the IoT device (step 1), verifies the device identity using the Manufacturer Identification smart contract (step 2), and fetchs the MUD file reference from the Device Registry contract (step 3).
The MUD file reference is then used to fetch the actual MUD file from an off-chain database (step 4), which is  used by the SDN controller to configure access control policies for the deployed IoT device.
By adopting our blockchain-based platform transparency and auditability of this information increases the reliability and security.

\lstset{breaklines=true, basicstyle=\scriptsize, morekeywords={contract, event, modifier, constructor, function, address, string, bytes32, indexed, public, memory, require}}
\begin{lstlisting}[frame=single, caption={Identity management contract}, label=contract-id]
contract IdentificationAuthorithy {

  address public owner;
  string public ownerName;
  string public ownerCertificate;
    
  event CertificateIssued (
    address indexed owner,
    string  indexed ownerName,
    string  indexed subjectName,
    string  subjectCertificate);    

  constructor(address _owner, 
      string memory _ownerName, 
      string memory _ownerCertificate) public {
    owner = _owner;
    ownerName = _ownerName;
    ownerCertificate = _ownerCertificate;
  }

  modifier onlyOwner() { 
    require(msg.sender == owner);  _; 
  }
    
  function issueCertificate(
      string calldata _subjectName, 
      string calldata _subjectCertificate)
      external onlyOwner {
    emit CertificateIssued(msg.sender, ownerName, 
    _subjectName, _subjectCertificate); 
  }
    
  function revokeCertificate (...) {
  
  }
  
\end{lstlisting}

The manufacturer identification and device registry smart contracts were implemented using the EVM (see Section \ref{sec:blockchain-background}) in a private blockchain deployment for evalutation purposes.
The manufacturer identification contract is an instance of the general purpose identification authority contract described in Listing 1, where some implementation details were omitted due to space restrictions.
This same contract can be instantiated by the EU Identification Authority, National Certification Authority of each Member State, and Device Manufacturers as introduced in Figure \ref{fig:identification}.
The use of a smart contract for identity management is a clear advantage since it is similar to a log-based Public Key Infrastructure (PKI), which allows transparency with respect to the issuance of identity assurance certificates \cite{Al-Bassam:2017}.

Listing 2 shows the device registry contract.
This contract is created by the manufacturer for each new device type and defines a \textit{registerFile} function that can only be invoked by the manufacturer or a assessment body also specified by the manufacturer.
All the history of the device registry is kept in the blockchain since it can only be modified by function invocations done through transactions.
Details about the vulnerability disclosure functions and functions to collect feedback from end-users to establish a device reputation were omitted due to space restrictions.

\lstset{breaklines=true, basicstyle=\scriptsize, morekeywords={contract, event, modifier, constructor, function, address, string, bytes32, indexed, public, memory, require}}
\begin{lstlisting}[frame=single, caption={Device registry contract}, label=contract-registry]
contract DeviceRegistry {
 
  address public manufacturer;
  string public manufacturerName;
  address public manufacturerIdContract;
  string public deviceId;
  address public assessmentBody;
  string public assessmentBodyName;
  address public assessmentBodyIdContract;
    
  event RegisterFile (
    address sender,
    string indexed deviceId,
    string indexed fileType,
    string indexed fileLocation,
    bytes32 fileHash);

    constructor(address _manufacturer,
        string memory _manufacturerName, 
        address _manufacturerIdContract,
        string memory _deviceId) public {
      manufacturer = _manufacturer;
      assessmentBody = _manufacturer;
      manufacturerName = _manufacturerName;
      manufacturerIdContract = _manufacturerIdContract;
      deviceId = _deviceId; }

  modifier onlyManufacturer() {
    require(msg.sender == manufacturer);  _; }
  modifier onlyAssessmentBody() { 
    require(msg.sender == assessmentBody);  _; }
    
  function setAssessmentBody(address _assessmentBody,
      string calldata _assessmentBodyName, 
      address _assessmentBodyIdContract) external onlyManufacturer {
    assessmentBody = _assessmentBody;
    assessmentBodyName = _assessmentBodyName;
    assessmentBodyIdContract = _assessmentBodyIdContract;
  }

  function registerFile(string calldata fileType,
      string calldata fileLocation,
      bytes32 fileHash
      ) external onlyManufacturer onlyAssessmentBody {
    emit RegisterFile(msg.sender, deviceId, fileType, fileLocation, fileHash); }
}
\end{lstlisting}

Finally, Listing \ref{mudexample} shows a MUD file example for a temperature sensor that was stored in an off-chain database and registered in the device registry smart contract using the $registerFile$ function (see Listing 2).
According to the MUD Model \cite{mudrfc}, the \textit{ietf-mud} container includes different parameters, such as the URL where the MUD file can be downloaded, or the date of the last update. It also contains the name of the access control lists, which are specified in the \textit{ietf-access-control-list} container. In this case, the ACL \textit{mud-37547-v6to} restricts the access to the device and the ACL \textit{mud-37547-v6fr} restricts the access
from the device. Such access control lists are intended to allow the communication with devices from the same manufacturer (manufacturerA), through the ports 33 and 12 and using the UDP protocol (coded as 17).

\lstset{language=JAVA, breaklines=true, basicstyle=\scriptsize}
\begin{lstlisting}[frame=single, caption="MUD File Example", label=mudexample]
{"ietf-mud:mud": {
  "mud-version": 1,
  "mud-url": "https://host1/model1",
  "last-update": "2019-05-16T09:03:46+00:00",
  "cache-validity": 48,
  "is-supported": true,
  "systeminfo": "Temperature sensor",
  "mfg-name": "manufacturerA",
  "documentation": "https://www.documentation.org",
  "model-name": "model1",
  "from-device-policy": {
    "access-lists": {
      "access-list": [{"name": "mud-37547-v6fr"}]}},
    "to-device-policy": {
      "access-lists": {"access-list": [{ "name": "mud-37547-v6to" }]}}},
  "ietf-access-control-list:acls": { "acl": [{
    "name": "mud-37547-v6fr",
    "type": "ipv6-acl-type",
    "aces": { "ace": [{
      "name": "myman0-frdev",
      "matches": { "ietf-mud:mud": {
        "same-manufacturer": "manufacturerA" },
        "ipv6": { "protocol": 17 },
        "udp": {
          "destination-port": {
            "operator": "eq","port": 33},
          "source-port": {
            "operator": "eq","port": 12}}},
          "actions": {"forwarding": "accept" }}]}}]}}
\end{lstlisting}

\section{Conclusions and Future Work}
\label{sec:conclusions}

Cybersecurity certification represents a cornerstone in the landscape at EU level. Although the definition of a certification framework sets out significant challenges, its realization is expected to increase users' trust in the digital era. One of the most significant needs is due to the lack of a common platform to share cybersecurity information in a secure and transparent way. We believe that our proposed blockchain-based approach is intended to serve as a referece point where manufacturers, end users and organizations could get access and update the certification information of IoT devices.
As future work we propose to investigate two improvements to the approach proposed in this paper.
Firstly, we plan to implement the proposed architecture using Hyperledger following a hierarchical sharding approach where the blockchain is not anymore a centralized data structure but is split in multiple layers where each layer above acts as a synchronization point.
Secondly, we plan to integrate this approach to our previous work where data usage control policies were deployed and evaluated in smart contracts \cite{neisse:ares2017}, with an additional feature where devices can verify policy decisions using a lightweight blockchain synchronization protocols (e.g. Ethereum light node protocol \cite{eth-light-node}) without relying on external entities.
Our goal with these two improvements is to provide a fully decentralized approach for cybersecurity management of IoT devices, which after evaluation could be generalized to any type of device, product, or service in line with the current EU legislative documents.

\section{Acknowledgements}
This work has been partially funded by the European Commission through the H2020-779852 IoTCrawler project, the H2020-780139 SerIoT project, the FPU-16/03305 research contract of the Ministry of Education and Professional Training of Spain, the project PERSEIDES (TIN2017-86885-R) granted by the Ministry of Economy and Competitiveness of Spain and CHIST-ERA PCIN-2016-010.

\bibliographystyle{IEEEtran}
\bibliography{securityiotblockchain}
\end{document}